\input amssym.def
\input amssym.tex
\hsize=16.0 truecm
\vsize=24.0 truecm
\baselineskip=7truemm plus 2truemm   
\parskip=5pt plus 2pt
%
%

\font \bigfonts=cmr9 scaled\magstep2
%
\def\leaderfill{\leaders\hbox to 1em{\hss.\hss}\hfill}
\def\hb#1{\hbox to .6cm{\hss#1}}
\font\gr=cmmib10
\def\alphabf{{\hbox{\gr\char11}}}
\def\d{ \displaystyle}
\def\ltende{\longrightarrow}
\def\tende{\rightarrow}

\def\abs#1{\vert #1 \vert}

\def\bye{\par\vfill\supereject\end}
\def\bl{\hfil\break}

\def\lb{\langle}
\def\rb{\rangle}

\def\om{\omega}
\def\wp{W_{+}}
\def\wm{W_{-}}
%
\def\abst#1{\centerline{
\hbox {\hsize=11.0 truecm
\baselineskip=4.5truemm minus 1truemm 
\vbox to 1.0in{
\font \picc=cmr8
\noindent
\picc
 #1}}}\bl}
\def\abst1#1{\centerline{
\hbox {\hsize=13.5 truecm
\baselineskip=5.0truemm minus 1truemm 
\vbox to 1.0in{
\font \piccc=cmr9
\noindent
\piccc
 #1}}}\bl}

\font \itp=cmti8

\font \bigbf=cmbx10 scaled\magstep2
\def\pfp{{P_{FP}}}
\def\pt{{\tilde P}}
\def\ref#1#2#3#4#5#6{\item{[#1]} #2, {\it #3 \/}{\bf #4}, #5 (#6).} 
%
\ \ \hbox{\hskip 11.4truecm}DFBO/95/27\vskip 1.1truecm\noindent
\centerline{\bigbf Finite Size Scaling Analysis of Biased Diffusion on 
Fractals}
\vskip 1.9 truecm\noindent
\centerline{Giovanni Sartoni$^{1*}$ and Attilio 
L. Stella$^{2\dag}$}\vskip 0.6truecm\noindent
\centerline{\itp
$^1$ Dipartimento di Fisica and Sezione INFN, Universit\`a di 
Bologna, I-40126 Bologna Italy\/}\par\noindent
\centerline{\itp
$^2$ INFM--Dipartimento di Fisica and Sezione INFN, 
Universit\`a di Padova, I-35131 Padova Italy\/}\bl\vskip 0.4truecm\noindent
\vfootnote{*}{Email: sartoni@bo.infn.it}
\vfootnote{\dag}{Email: stella@bo.infn.it}

\centerline{(May 13, 1996)}\vskip 1.5truecm\noindent
\centerline{\bigfonts Abstract}\vskip 1.0truecm\noindent
\abst1{Diffusion on a T fractal lattice under the influence of topological 
biasing fields is studied by finite size scaling methods. This allows to avoid 
proliferation and singularities which would arise in a renormalization group 
approach on infinite system as a consequence of logarithmic diffusion. 
Within the scheme, logarithmic diffusion is proved on the basis of an 
analysis of various temporal scales such as first passage time 
moments and survival probability characteristic time. This confirms and puts on 
firmer basis previous renormalization group results. A careful study of the 
asymptotic occupation probabilities of different kinds of lattice points allows 
to elucidate the mechanism of trapping into dangling ends, which is responsible 
 of the logarithmic time dependence of average displacement.
\vskip 6.2truemm\noindent
\hbox{Pacs numbers: 64.60.Ak, 05.40.+j, 05.60.+w.}\bl
}
\vfill\supereject
{\bf 1 Introduction}\bl
A particle diffusing under the influence 
 of a random biasing field can
  exhibit slow diffusion, with average displacement, $R$, growing in time,  
 $t$, as some power of $\ln t$ [1-7]. A remarkable exact demonstration of logarithmic  
 diffusion for a particle hopping on a one-dimensional chain and subject at  
 each site to an independent random bias is due to Sinai [7]. There is also
 evidence, based upon numerical results, that logarithmic diffusion 
replaces the anomalous power 
 law one, when particles diffuse on fractal structures, like the incipient infinite  
 cluster of percolation, and are
 subject to a uniform biasing field [8,9]. In this 
 case, though the field is not random, the complexity of the fractal structure conspires 
 with the former in determining possible localization effects, e.g., by pushing and 
 trapping the moving particle into dangling ends. 
So far, this phenomenon has been modelized, in most cases, by diffusion along comblike 
 structures with teeth of variable length, where a bias pushes the particle towards 
 the tips [10]. Despite the efforts produced, a comprehensive and 
 satisfactory mathematical description and a general physical understanding of 
 the phenomenon of logarithmic 
diffusion are still lacking. Present insight 
 is mostly limited to the  
 example of a particle diffusing in 1D and subject to 
a random bias [11]. The case of diffusion on fractal structures under constant 
bias, which is our concern here, is more problematic. Comblike models of this 
latter type of diffusion are again basically 
 one-dimensional, and rather far from the complexity of self-similar 
structures. The accepted picture of bias induced logarithmic  
 diffusion on a fractal, through the mechanism of trapping into dangling ends,  
 is intuitively appealing but rather vague. So, it is worth 
to further elucidate and test it, in order 
to have a more reliable description of such a process.\bl
Recently a study of such trapping mechanism has been 
 accomplished by a dynamical renormalization group (RG)  analysis of diffusion
on deterministic 
 fractal lattices under the influence of a topological bias [12].
 Deterministic self-similar structures,
 though being a mathematical schematization, retain various fundamental features
 of real random systems, with the advantage of allowing analytical solutions 
 of random walk processes [13-15]. Furthermore, the RG has demonstrated to be a powerful 
 tool in treating dynamics on self-similar structures [14,15], 
also in the very peculiar  singular 
conditions implied by logarithmic diffusion [12]. Indeed, one should notice 
 that logarithmic time behaviour implies a singular structure for the
 RG transformation. A standard dynamical RG yields the time rescaling factor 
 $l^z$, under length rescaling $l$, which is consistent with power law anomalous 
 diffusion, $R\sim t^{1/z}$. So, logarithmic diffusion clearly corresponds to the 
 $z\tende +\infty$ limit, causing $l^z \tende \infty$ as well, and thus a 
 singularity in the RG transformation. As claimed in ref.[12], the control of a 
singular RG transformation should be the key to extract the correct logarithmic 
time dependence of $R$. On the other end, a singular RG 
transformation can be useful
 only if the RG is carried out exactly [12]. Such a requirement 
and the consequent 
 need of handling extremely strong singularities reveals a very
 hard task, implying, by necessity, some ad hoc simplifications and
 assumptions in the RG approach [12].\bl
In alternative to RG, finite size scaling (FSS) analysis of non-equilibrium 
dynamical processes [16]
 can be exploited for the study of power-law anomalous diffusion. Recently a
 remarkably efficient  and straightforward approach to diffusion on finite self-similar
trees and hierarchical combs has been proposed [17]. In this approach, through 
an exact 
 decimation of the master equations and consequent rescalings, the
 FSS of first-passage time (FPT) moments and survival probability (SP) 
 can be studied.\bl
A FSS based phenomenological RG study of anomalous diffusion
offers several advantages with respect to conventional RG strategies. 
 In particular one does not need to care about proliferation problems in the dynamical 
equations. Thus, one can legitimately wonder whether a similar approach could 
be used to confirm and elucidate 
logarithmic anomalous diffusion on fractal structures of the 
type considered in ref.[12].\bl
In the present paper we apply the methods of ref.[17] to address
 diffusion on deterministic T-like finite fractal trees in two dimensions, 
when a topological bias field is acting. We thus generalize these 
finite size scaling methods [17] to a system exhibiting singular 
dynamical scaling. At a first sight,  this singular scaling could 
 seem to prevent any effective analysis altogether. 
To the contrary, we are able to derive the 
 FSS behaviour of FPT moments, SP, first-passage probability (FPP) and probability  
 of being at a certain site at time $t$ (the site occupation 
 probability). All these quantities turn out to be controlled by the same time scale, $\d
 t^* \sim e^{\alpha L}$, $L$ being the linear size of the finite lattice. This 
 confirms that, in the infinite system, 
diffusion is logarithmically localized within a distance $R\sim \ln t$, 
 after a time $t$, as already conjectured [12]. Furthermore, as discussed below,
 the existence of a single time scale, $t^*$, characterizing 
both 
 the  fixed observation-point ensemble (yielding the average over the first 
passage time at a fixed given site, i.e. the FPT moments) and the fixed 
observation-time one (concerning the average over the walker displacements 
at time $t$, $\lb 
R(t)\rb$ ), supplies concrete 
 support to the intuitive hypothesis that bias induced trapping
 into dangling ends is directly responsible of logarithmically slow 
diffusion on self-similar 
 structures.\bl
As a further point of interest, our results show
 that the presence of a singularity in RG 
 transformations at the fixed point does not prevent the FSS hypothesis to hold 
 in this case. This was not a priori obvious in such a circumstance.\bl 
As anticipated above, from a methodological viewpoint, 
the strategy of studying
 logarithmic diffusion by FSS allows to avoid some of the difficulties 
 involved in renormalising the infinite model. Since finite size analysis  
 requires to consider the infinite time limit of physical quantities while 
 keeping linear dimensions finite, asymptotic time behaviours are 
extracted
 while divergences, induced by the singular time rescaling, still have to 
develop fully. Hence no ad hoc assumptions are needed to handle strong 
 divergences, as it is the case in infinite systems [12].\bl
This article is organised as follows. In the next section we introduce the 
model of biased diffusion on deterministic T fractal, and 
 discuss  quantities 
relevant to the description of the random walk process. In the third 
section the 
decimation technique of the finite set of master equations is developed 
and applied to our 
model. Phenomenological RG transformations of FPP and SP are then derived and 
the scalings of FPT 
and characteristic time are calculated. The methods
 of section 3 are reconsidered in  
section 4 and used to extract the long time behaviour of occupation
 probabilities. 
Several variants of the model, differing in initial and/or boundary conditions  
are also examined in section 4, to show that these different conditions 
do not influence the main 
dynamical scaling properties. Finally section 5 is devoted
 to discussion and 
conclusions.\par 
 {\bf 2 Deterministic fractal model of biased diffusion}\bl
Diffusion on random ramified self-similar structures can be fruitfully 
investigated 
by the study of random walks on deterministic fractal lattices [13-15]. Often 
these offer the advantage of being simple enough to allow analytical treatment.
 For definiteness we will focus on a particular realization of 
 our model, the so-called T fractal (see Fig.2.1), a hierarchical tree in 2D, 
 whose fractal dimension is ${\bar d}=\ln 3/\ln 2$ [18]. On such a lattice at 
 some finite generation order $n$, a particle diffuses  
 by hopping between nearest neighbour (nn) sites. The probability,
 $P_{i} (t)$, for the particle to be at site $i$ at time $t$, obeys the master 
 equation:
$$\d P_{i}(t+\tau )=P_{i}(t)+\sum_{j\ne i}\left[ W_{ij} P_{j}(t)-W_{ji} 
P_{i}(t)\right]\ \ \ ,\eqno(2.1)$$
where the sum on the right hand side runs over nn, $j$, of site $i$, 
and $W_{ij}$ is the probability that a particle hops 
from $j$ to $i$ in time $\tau$. Here it is assumed
 that $W_{ij}=\wp$ or $\wm$, according to whether one goes from site $j$ to $i$ 
 following the bond arrow or not, respectively. Arrows are 
all pointing in the direction of increasing ``chemical'' distance 
from the lattice's leftmost site (site 1 of 
 Fig.2.1). Thus the following consistency relations have to be satisfied: 
 $2\wp +\wm \le 1$, $W_{\pm} \le 1$. Moreover, we assume 
 $W=\wm /\wp \le 1$, so that the bias tends to drive the particle towards 
 longer chemical distance from site 1. This is an example of topological bias 
field, namely a field having a particular direction in the topological space where 
the hopping process takes place. This is different from the so called Euclidean 
bias, which is uniformly directed in the Euclidean space onto which
 the lattice 
in embedded [13].\bl 
We choose site 1 to be the starting point for particle motion: so, the initial 
condition of eq.(2.1) is $P_{i}(0)=\delta_{i,1}$.
 We also impose the rightmost end-tip site of the lattice, $i_s$, (e.g. 
number 4 or number
 10 of Fig.2.1a or b, respectively) to be a fully absorbing sink, i.e. 
a position from which the particle can not hop back to other sites of the 
T fractal ($W_{ji_s}=0\quad\forall j$).\bl
Due to the absorbing site, the SP has an exponential decay whose 
characteristic time is
 related to the linear size of the lattice. At generation 
$n$, a T fractal has linear size $L=2^{n+1}$. In standard anomalous diffusion 
processes,
 where the average displacement, $R$, scales as $R\sim t^{1/z}$, 
the characteristic
 time, $t^*$, is expected to scale with $L$ as $t^* \sim L^z$. Relying 
 on previous results for the same model without sink and in the infinite size 
limit [12], we expect in the 
 present case a logarithmic localization of motion too, thus a scaling $t^* \sim
 e^{\alpha L}$, with $\alpha >0$.\bl
Moreover, since $i_s$ is absorbing, its occupation probability,
 $P_{i_s}
 (t)$, coincides exactly with the first-passage probability, $P_{FP}(t)$,
 at time $t$ for that
 site. $P_{i_s}$ is also closely related to the survival probability, $S(t)$, 
namely the
 probability that the particle has not yet reached the output (sink) site at time $t$. 
 Indeed,
$$\d S(t)=1- \sum_{k=0}^m \pfp (k\tau)\ \ , \hbox{(with $t=m\tau$).}\eqno(2.2)$$
Putting $\tau=1$, and applying the time discrete Laplace transform [14,19],
$$\d \pt_{i}^0 (\om_0 )=\sum_{m=0}^\infty P_i (m)(1+\om_0 )^{-1-m} \quad ,\eqno
(2.3)$$
to eq.(2.1), we get:
$$\d \left[\alpha (i,\om)+\sum_{j\ne i} {W_{ji}\over\wp} \right] \pt_i 
(\om)=\sum_{j\ne i} {W_{ij}\over \wp} \pt_j (\om) +\delta_{i,1} \ \ \ ,
\eqno(2.4)$$
where $\om =\om_0 /\wp$, $\pt_{k}^0 (\om_0 )=\pt_k (\om)/\wp$, and the Kronecker's 
 delta reflects the initial condition $P_i (0)=\delta_{i,1}$. According to eq.(2.1) 
 one should have $\alpha (i,\om)=\om \ \ \ \forall i$, in eq.(2.4). However,
 we already introduce these coefficients in view of the fact that 
sites with different coordination are 
 inequivalent under lattice decimation and consequent rescaling, 
as we shall see later. 
 It will turn out that $\alpha (i,\om )=\alpha_1 (\om ),\ \alpha_3 (\om )$ or $\alpha_0 
 (\om )$, according to whether $i$ has coordination 1, 3 or coincides with site 1, 
 respectively. The only parameters entering eq.(2.4) will thus be ${\alphabf}=
 (\alpha_0 ,\alpha_1 ,\alpha_3 )$, and $W$.\bl
When the sink site $i_s$ is considered, we see 
immediately that Laplace-transforming 
  $P_{i_s} (m)=\pfp (m)$, provides the generating function for the FPP, indeed:
 $$\d (\om_0 +1){\partial \pt_{FP}^0 (\om_0 )\over\partial\om_0}
 \Big\arrowvert_{\om_0 =0} =- \sum_{m=0}^\infty 
 (m+1) \pfp (m)=- \lb m\rb -1\ \ ,\eqno(2.5)$$
$\lb m\rb =\lb t\rb$ is the average FPT. Higher moments of FPT come from higher 
 derivatives of $\pt_{FP}^0$. E.g., by 
deriving $\pt_{FP}^0$ two times one finds:
$$\d (\om_0 +1){\partial\over \partial\om_0} \left( (\om_0 +1) {\partial\over
\partial 
\om_0} \pt_{FP}^0 (\om_0 )\right) \Big\arrowvert_{\om_0 =0} = \sum_{m=0}^\infty (m+1)(
m+1) \pfp (m)=\lb m^2 \rb +2\lb m\rb +1 \ .\eqno(2.6)$$
Clearly most informations about the diffusion process are contained in 
 $\pt_{i}^0 (\om_0 )$, or equivalently in $\pt_i (\om )$. Thus, the
 analysis of these quantities  will be our main concern below.\par
{\bf 3 First-passage time and characteristic time scaling}\bl
The first-passage probability for the T fractal with an absorbing point at the 
 right end tip, can be calculated by successive decimations of the 
Laplace-transformed 
 master equation. Consider the zeroth-order tree of Fig.2.1a, 
with site 1 being  
 the input site and site 4 the output one. The explicit form of eq.(2.4) reads:
$$\d \cases {(\alpha_0 (\om )+1)\pt_1 =W\pt_2 +1\cr 
\ \cr
\d (\alpha_3 (\om )+2+W) \pt_2 =\pt_1 +W \pt_3\cr
\ \cr
\d (\alpha_1 (\om )+W) \pt_3 =\pt_2 \cr
\ \cr
\left( \om +{1\over \wp}\right)\pt_4 = \pt_2\cr}
 . \eqno(3.1)$$
Solving this system for $\pt_4$ yields:
$$\d \pt_4 (\om ) ={\alpha_1 (\om )+W \over (\om +{1\over \wp }) 
 \left\{ \left( \alpha_0 (\om )+1\right) \left[
 \left( \alpha_3 (\om )+2+W \right)  \left( \alpha_1 (\om )
+W \right) -W \right] 
- \left( \alpha_1 (\om ) +W \right) W \right\} } \ . \eqno(3.2)$$
Furthermore, the Laplace-transformed master equation system 
for the first order tree of Fig.2.1b is:
$$\d \cases {(\alpha_0 (\om )+1) \pt_{1}  =W\pt_{2} +1 \cr
\ \cr
\d (\alpha_1 (\om )+W) \pt_{i} = \pt_{j} \cr
\ \cr
\d (\alpha_3 (\om )+2+W) \pt_{i} = \pt_{k} +W \d\sum_{j\ne k} \pt_{j} \cr
\ \cr
\d (\alpha_3 (\om )+2+W) \pt_{8} = \pt_{4} +W \pt_{9}\cr
\ \cr
\left( \om +{1\over \wp} \right) \pt_{10} =\pt_{8} \cr}\ \ \ ,\eqno(3.3)$$
Decimating the last generated sites, namely 2, 3, 5, 6, 8, and 9, 
and eliminating 
 the respective $\pt_i$'s from system (3.3), we go back to a lattice which is 
 equivalent to the first order one, after 
rescaling lengths by a factor 2. After suitable 
 redefinitions and site index reordering, the system for the surviving 
$\pt$'s can be reduced to a
 form analogous to (3.1),
$$\d \cases {(\alpha_{0}^{(1)} (\om )+1) \pt_{1}^{(1)} =W^{(1)}\pt_{2}^{(1)} 
+1 \cr
\ \cr
\d (\alpha_{3}^{(1)} (\om )+2+W^{(1)} ) \pt_{2}^{(1)} =\pt_{1}^{(1)} +W^{(1)}
\pt_{3}^{(1)} \cr
\ \cr
\d (\alpha_{1}^{(1)}(\om) +W^{(1)} )\pt_{3}^{(1)} =\pt_{2}^{(1)} \cr
\ \cr
 \left( \om +{1\over \wp}\right) \pt_4 = \pt_{2}^{(1)}\cr} \ \ \ ,\eqno(3.4)$$
with the following identifications:
$$W^{(1)} =W^2 \eqno(3.5a)$$
$$\d \alpha_{0}^{(1)} (\om )={(\alpha_0 (\om )+1) \left[ (\alpha_1 (\om )+W)
(\alpha_3 (\om )+2+W)-W \right] \over \alpha_1 (\om )+W} -1-W \eqno(3.5b)$$
$$\d \alpha_{1}^{(1)} (\om )= (\alpha_1 (\om )+W)(\alpha_3 (\om )+2+W) -2W- W^2
\eqno(3.5c)$$
$$\d \alpha_{3}^{(1)} (\om )= {(\alpha_3 (\om )+2+W) \left[ (\alpha_1 (\om )+W)
( \alpha_3 (\om )+2+W)-W \right] \over \alpha_1 (\om )+W } -2-3W- W^2 \eqno(3.5d)$$
$$\d \pt_{i'}^{(1)} (\om )= {\alpha_1 (\om )+W \over (\alpha_1 (\om )+W) (\alpha_3 
(\om )+2+W) -W } \pt_{i} \ \ \ ,\eqno(3.5e)$$
with $i'=1$ if $i=1$, $i'=2$ if $i=4$, and so on. The apex (1) in front of all
left hand quantities indicates that they have been rescaled once after a 
decimation. A few remarks about these RG transformations: in first place, 
 quantities relative to the sink site do not rescale, 
as it appears from the last 
 of eqs.(3.4). In addition one should notice that different $\alpha_{i}^{(1)}$'s
 undergo different transformations, the dependence on the original 
$\alpha_i$'s being nonlinear. This 
is a standard
 memory effect associated with dynamical coarse graining. Now the necessity of 
 introducing different $\alpha$-coefficients becomes clear.\bl
Starting with an $n$th-order lattice, 
we perform $n$ successive decimations
 and rescalings in such a way 
to arrive to the zeroth-order system. Then, solving the latter for 
 $\pt_4$, the FPP generating function, we find:
$$\d \pt_{FP} (\om )= \d{ \alpha_{1}^{(n)} (\om )+ W^{(n)} \over \left( \om +{1\over
 \wp} \right) \Delta^{(n)}} \eqno(3.6)$$
$$ \Delta^{(n)}=  ( \alpha_{0}^{(n)} (\om ) +1) \left[ 
 ( \alpha_{3}^{(n)} (\om )+2+ W^{(n)} )
 ( \alpha_{1}^{(n)} (\om )+ W^{(n)} ) -W^{(n)} \right] 
 - W^{(n)} ( \alpha_{1}^{(n)} (\om )+W^{(n)} ) \ .$$
Moments of the first passage time are obtained by  the series expansion of  
 $\pt_{FP}$. Indeed assuming that $\pt_{FP}$ is expandable around $\om=0$ 
(which is justified for a finite system) i.e.
$$ \pt_{FP} (\om )= p_0 +p_1 \om +p_2 \om^2 +O(\om^3 ) \ \ \ ,\eqno(3.7)$$
we see immediately from (2.5-6) that $\d p_1= -\lb t \rb -1$  and $\d p_2 =
\lb t^2 \rb +2 \lb t \rb +1$ (notice that, since $\d {\partial \pt_{FP}^{0}
 (\om_0 ) \over \partial \om_0 } = {1\over \wp } {\partial \pt_{FP} 
(\om_0 /\wp ) \over \partial \om_0} 
 = {\partial \pt_{FP} (\om ) \over \partial \om }$, $p_i$ are related 
to FPT moments without 
 any further rescaling). Thus, once the relation between the $p_i$'s of
 two successive T Fractal orders is known, one can derive  the scaling of 
 $\lb t \rb$. Since we are interested only in the average FPT we focus on $p_1$ 
 rescaling. To this aim we have to expand the right hand side of (3.6) 
and inherent 
 quantities up to first order in $\om$ about $\om=0$.\bl
To calculate ${\alphabf}^{(n)} (\om )$ we first consider its series expansion
in powers of $\om $
 about $\om =0$. Writing $\d \alpha_{i}^{(n)}(\om )=\sum_{k}^{\infty} 
\alpha_{i}^{(n),k} \om^k $, and considering (3.5b-d), we are able 
to derive the
  relation between ${\alphabf}^{(n),1}$ and ${\alphabf}^{(n-1),1}$, which are 
needed to calculate the scaling of $p_1$. By 
 keeping everywhere only leading order in $1/W^{(n)}$ ($W^{(n)} =W^{2^n} 
=W^{L/2}$),
 in the explicit relations, we get:
$$\d \cases {\alpha_{0}^{(n),1} \cong \d { (1+W) \alpha_{1}^{(n-1),1} 
\over W^{(n-1)} } 
\cong \d { (1+W) 2^{n-1} \over W^{2^{n-1}} }\cr
\ \cr
\alpha_{1}^{(n),1} \cong  (1+W) 2 \alpha_{1}^{(n-1),1} \cong (1+W)2^n \cr
\ \cr
\alpha_{3}^{(n),1} \cong \d  { (1+W) 2 \alpha_{1}^{(n-1),1} \over W^{(n-1)} } 
\cong \d 
{(1+W) 2^n \over W^{2^{n-1}} }\cr} \ \ \ . \eqno(3.8)$$
Disregarded terms in (3.8) are $\d o( W^{2^{n-2}})$, which 
 is negligible for $n\gg 1$, since $W<1$ in the biased case. Expanding (3.6) 
 consistently with the approximation leading to (3.8) then yields:
$$\d \pt_{FP} (\om ) \cong {\wp \over W^{(n)}} \left[ W^{(n)} -\om 
\alpha_{1}^{(n),1} +O(\om^2) \right] \ \ \ . \eqno(3.9)$$
So $\lb t \rb$ scales as
$$\d \lb t \rb^{(n)} +1 \cong { 2\over W^{(n-1)} }\left( \lb t\rb^{(n-1)} +1
\right) \ \ \ ,\eqno(3.10)$$
and finally we have:
$$\d \lb t\rb^{(n)} \sim { 2^n \over W^{2^{n}} }
={ L \over 2 W^{L/2} } ={L \over 2}\d e^{\abs{\ln W} L/2} \ \ . \eqno(3.11)$$
The survival probability, $S(m)$, defined in (2.2) decays exponentially in $m$ 
 as $\d S(m) \propto \exp (-m/ t_{n}^*)$, at long times, for any finite 
$n$th-order  T fractal, due to the absorbing boundary conditions. 
So, we want to investigate the relation between the characteristic time, 
 $t_{n}^*$, and the FPT. $P_{FP} (m)$ decays also exponentially, because it is 
 just the discrete derivative of $S(m)$: $\d P_{FP} (m) {\buildrel  
{\scriptscriptstyle m\tende +\infty}\over \propto}
 \left( e^{1/t_{n}^*} -1 \right) e^{-m/ t_{n}^* }$. This implies that 
 $\pt_{FP} (\om )$ has a simple pole in $\d \om_c ={1\over \wp } \left( e^{-1/
 t_{n}^* }-1 \right)$ in the limit $\om \tende 0$, which corresponds to the 
 long time behaviour of $P_{FP} (m)$.\bl 
The characteristic time grows with the T fractal size $n$, $\d t_n^* 
{\buildrel  {\scriptscriptstyle n\tende +\infty}\over\ltende}+\infty$
; hence $\d \om_c {\buildrel  {
\scriptscriptstyle n\tende +\infty } \over\ltende} 0^-$ 
and can be located by looking for the simple pole of
 $\pt_{FP}$ in the limit $\om \tende 0$ [20]. Expanding 
numerator and denominator 
 up to order $\om$ on the r.h.s of (3.6), and exploiting 
(3.8) for the expansion of
 ${\alphabf}^{(n)} (\om )$, we eventually arrive to:
$$\d \pt_{FP} (\om )\simeq {\alpha_{1}^{(n),1} \om +W^{(n)} \over \left( \om +
{1\over \wp} \right) \left( \om +{ (1+W) W^{(n)} \over 2 \alpha_{1}^{(n),1} }
\right) } \ \ . \eqno(3.12)$$
The first simple pole $\om_c = -{1 \over \wp}$ is not related  to long time 
 behaviour, since $- {1 \over \wp} \le -1$. The second simple pole is $\d
 \om_c \simeq  -{W^{(n)} \over 2 \alpha_{1}^{(n),1} } \simeq - {W^{2^{n}} \over
 2(1+W) 2^n } = - {W^{L/2} \over (1+W)L}$. Since $\d \om_c {\buildrel 
{\scriptscriptstyle n \tende +\infty}\over \cong } { -1 \over \wp t_{n}^* }$, 
then
$$\d t_{n}^{*} \sim {2^{n+1} \over W^{2^{n}} } \sim 
{L \over W^{L/2} }\ \ .\eqno(3.13)$$
So diffusion has a characteristic time scale growing exponentially with the  
 system size, and the dynamical exponent, $z$, diverges as expected. Moreover 
 $t_{n}^* $ scales like $\lb t \rb^{(n)}$, and, consistently with
 infinite  lattice
 result for the average displacement [12], the typical distance, 
$L$, travelled  in  time 
 $t$ is $L\sim \ln t$.\bl
All the above results provide evidence that dynamics is controlled by a single
 time scale determined by the trapping mechanism into dangling ends. In truly 
 infinite fractal structures [12] dangling ends  of all length scales exist, 
 thus dynamics is logarithmically localized.\bl
The existence of a single characteristic time suggests also that fixed time
 and fixed point of observation ensembles are substantially equivalent in this model,
 in accordance to what happens in most models of power-law 
anomalous diffusion on self-similar lattices [13-17].\bl
Finally, two important remarks concerning the approximate calculations are in 
order. 
 In first place, due to approximation to 
$o(W^{2^{n-2}})$ introduced to obtain eqs.(3.8) 
 and consequently (3.9) and (3.12), results 
concerning $\lb t\rb$ and $t^*$ can not be 
 extended to the region $W\lesssim 1$, of cross-over
to the standard anomalous diffusion
 regime corresponding to $W=1$ [12,14]. Secondly, we assumed $\pt_{FP}
 (\om )$ to be
 expandable about $\om =0$ to get (3.9), while we have then seen it has a pole 
 $\om_c {\buildrel {\scriptscriptstyle n \tende +\infty }\over \ltende} 0^-$. 
However, since $\om_c$ is finite for any 
 finite $n$, no matter how large, this will cause no inconsistency, because for
 $\om$ close enough to $0$, an expansion will always be allowed. On the other 
 end, this feature implies that the limits $t\tende +\infty$ and 
$n\tende +\infty$ 
 (i.e. $L\tende \infty$) can not commute, because when $L\tende \infty$  
 a singularity of $\pt_{FP}$ sets exactly  at $\om =0$, thus results 
(3.9-11) do not 
 extend to that limit. This problem does not exist in principle for
 the asymptotic analysis of FPP: indeed, the derivation of (3.12,13) 
involves no analyticity 
 assumption for $\pt_{FP}$. However, the determination of the $n\tende +\infty$ 
limit  
 remains a serious task in its own due to the presence of strongly divergent 
 $1/W^{2^{n}}$ terms in RG equations [12].\par
{\bf 4 Long time behaviour of the occupation probabilities and 
trapping mechanism}\bl
The knowledge of the occupation probability, $P_{j}(t)$, of a generic 
site $j$ of the T fractal at its
 $n$th-order, at time $t$, supplies further information about the dynamics.
 $P_j (t)$ is also expected to decay exponentially at long times, 
and we are interested
 in its characteristic time, which should scale like $t^*$. Furthermore, since
 sites with different  
 coordination are dynamically inequivalent (look at different 
$\alpha_{i}$'s rescalings 
 in (3.5b-d)), such a feature should extend, in some way, to their occupation 
probabilities. When these show an asymptotic exponential decay, 
they should also 
have an amplitude factor dependent on site coordination and position, and 
on the bias parameter $W$ as well. The nature of such dependence may provide 
further insight into the diffusion mechanism.\bl
Asymptotic analysis applied to $P_{FP} (t)$ can be extended,
 with moderate additional effort, to get also the asymptotics of a 
generic $P_{i}(t)$.\bl
Consider again the $n$th-order T fractal with initial condition 
at site 1 and absorbing 
 boundary condition at the rightmost end-tip site. We perform $n$ 
successive decimations 
 of the master equations (2.4), eventually arriving  
 to a system, like (3.4), for the surviving $\pt^{(n)}$'s, rescaled $n-$times 
 according to (3.5). Now the system can be completely solved for the occupation 
 probabilities of surviving sites, yielding:
$$\d \pt_{3}^{(n)} (\om )={1\over \Delta^{(n)} } \eqno(4.1a)$$
$$\d \pt_{1}^{(n)} (\om )={ \left(\alpha_{1}^{(n)} (\om )+W^{(n)} \right) 
\left( \alpha_{3}^{(n)} (\om )+2+W^{(n)} \right) -W^{(n)} \over \Delta^{(n)} } \eqno(4.1b)$$ 
$$\d \pt_{2}^{(n)} (\om )= {\alpha_{1}^{(n)} (\om )+W^{(n)} \over \Delta^{(n)} }
\eqno(4.1c)$$
$$\d \Delta^{(n)} =( \alpha_{0}^{(n)} (\om )+1 )  
\left[ ( \alpha_{1}^{(n)} (\om )+W^{(n)} ) (
 \alpha_{3}^{(n)} (\om )+2+W^{(n)} ) -W^{(n)} \right] -W^{(n)} ( 
\alpha_{1}^{(n)} (\om )+W^{(n)} ) \ \ ,$$
and $\pt_{4}(\om )$ as in eq.(3.6). Using (3.5e), we then recover the original, 
 unrescaled, $\pt$'s as:
$$\d \pt_{i} (\om ) =\prod_{k=0}^{n-1} \left[ {( \alpha_{1}^{(k)}(\om )+
W^{(k)} )( \alpha_{3}^{(k)} (\om )+2+W^{(k)} ) -W^{(k)} \over \alpha_{1}^{(k)}
(\om )+ W^{(k)} } \right] \pt_{i'}^{(n)} (\om ) \ ,\eqno(4.2)$$
where $i'$ is the index that site $i$ takes after $n$ decimations. Again
  $\pt (\om )$ is a fraction and, being concerned with long time behaviours, we 
expand both numerator and denominator to leading order  
 in $\om$. In this way we expect to find the single pole 
corresponding to the 
 characteristic time, like already seen for $\pt_{FP} (\om )$. As 
additional task  here, the product preceding 
$\pt^{(n)}$ on the r.h.s. of (4.2), has to 
 be expanded. To leading order in $\om$, 
 this can be accomplished using the series expansions of ${\alphabf}^{(k)}(\om)$ 
about $\om =0$ and exploiting equations (3.8) for the coefficients 
$\alpha_{i}^{(k),1}$. Relations (3.8) are drawn within an accuracy of 
$o(W^{2^{n-2}} )$, which thus applies also to the results below. We shall 
omit the details of such cumbersome calculations, and give directly the 
asymptotic 
expression of occupation probabilities:
$$ \d P_1 (t) \approx {\left( 1-W^{2^n} \right) \over 1-W^2 } {W^{2^{n+1}} \over 
 4\  2^n } e^{-{t / t_{n}^* }} \eqno(4.3a)$$
$$\d P_4 (t) \approx {\left( 1-W^{2^n} \right) \over 1-W^2 } {W^{2^n} \over 4\  2^n}
 e^{-t/ t_{n}^*} \eqno(4.3b)$$
$$\d P_7 (t) \approx {\left( 1-W^{2^n} \right) \over 1-W^2 } {1 \over 2\  2^n} 
e^{-t/ t_{n}^*} \eqno(4.3c)$$
$$ \d P_{10} (t)\approx { \wp \over 4(1+W)} { W^{2^n} \over 2^n } e^{-{t /t_{n}^* }} 
\eqno(4.3d)$$
$$\d t_{n}^* \simeq {(1+W) \over \wp } {2^{n+1} \over W^{2^n} } \ \ \ .$$
For purposes which will be clear later on, sites have been labelled 
 with the indexes they have after $n-1$  decimations, consistently with
 Fig.2.1b. So 1 is the injection site, 10 is the absorbing one, 7 is 
 at the tip of the central branch  and 4 is at the lattice center (they  
correspond 
 to sites 1,4,3,2 of Fig.2.1a, respectively). Note that the characteristic 
time, 
 $t_{n}^*$, is the same we found in (3.13), for every site, while the 
amplitude  
 dependence on $W$ is rather varying from site to site.\bl
In order to have further terms of comparison
 it is worth deriving the asymptotic $P(t)$ form for a few other sites 
too, namely 
those which disappear due to the $n$th decimation, but have different 
chemical distances from both origin and sink. For definiteness we 
focus on 
sites 3,6 and 9 which are tips, and on 8 which has three nn (indexes refer 
always 
to Fig.2.1b). 
We suppose to arrest decimation at the $(n-1)$th stage, obtaining a 
lattice configuration 
 like Fig.2.1b. The site occupation probabilities to be calculated can be 
drawn by solving the corresponding $(n-1)-$times decimated master equation set
 for $\pt_i^{(n-1)} (\om)$, and using also results (4.1a-c). For 
example, site 6
 is at the same chemical distance from 1 as 7, then 6 and 7 are completely 
equivalent and their $P(t)$'s show identical long $t$ behaviour. 
On the other end , site  9,
 though equally far from the origin as 7, is influenced by the absorbing boundary  
 condition on the branch it belongs to and exhibits different features. Indeed:
$$\d \pt_{9}^{(n-1)} (\om )= { \pt_{8}^{(n-1)} \over \alpha_{1}^{(n-1)} (\om
 )+W^{(n-1)} } $$
(see e.g. system (3.3)), and since
$$ \d \pt_{8}^{(n-1)} (\om ) = { \alpha_{1}^{(n-1)} (\om )+W^{(n-1)} \over 
( \alpha_{1}^{(n-1)} (\om ) +W^{(n-1)} )(\alpha_{3}^{(n-1)}(\om  )+2+W^{(n-1)})
 -W^{(n-1)} } \pt_{4}^{(n-1)} =
\pt_{2}^{(n)} \ \ ,$$
we have
$$\d\pt_{9} (\om ) =\prod_{k=0}^{n-2} \left[ {( \alpha_{1}^{(k)} (\om )+
W^{(k)} ) ( \alpha_{3}^{(k)} (\om ) +2+W^{(k)} ) -W^{(k)} \over 
\alpha_{1}^{(k)}(\om )+W^{(k)} } \right] { \alpha_{1}^{(n)}(\om )+W^{(n)} \over \left( 
\alpha_{1}^{(n-1)}(\om )+W^{(n-1)} \right) \Delta^{(n)} } \ \ . \eqno(4.4)$$
Applying the above outlined asymptotic analysis to $\pt_{9}$, one finally 
arrives to
$$\d  P_9 (t) \approx { \left( 1- W^{2^{n-1}} \right) \over 1-W^2 } { W^{2^{n-1}}
\over 2\  2^n } e^{-t/ t_{n}^*} \ \ ; \eqno(4.5)$$
notice the $W^{2^{n-1}}$ extra factor, comparing to eq.(4.3c). 
In a similar way we calculate  occupation probabilities  of another tip site, number 3, 
which is closer to the bias origin, and another site with coordination 3, 
number 8,
$$\d  P_3 (t) \approx { \left( 1- W^{2^{n-1}} \right) \over 1-W^2 } {3 W^{2^{n}}
\over 16\  2^n } e^{-t/ t_{n}^*} \ \ , \eqno(4.6a)$$
$$\d  P_8 (t) \approx { \left( 1- W^{2^{n-1}} \right) \over 1-W^2 } { W^{2^{n}}
\over 4\  2^n } e^{-t/ t_{n}^*} \ \ . \eqno(4.6b)$$
The above results, (4.3,5,6), again point out how in a branch of linear
 size $L$,
 all occupation probabilities, as well as the FPP, decay with the same 
characteristic 
 time $\d t^* \sim e^{\abs{ \ln W} L/2}$. In an infinite structure with 
branches  
 of any length scale, the larger is the branch, the exponentially longer 
will be the time
 the particle spends in it, due to trapping effects caused by the bias. 
This will cause
 a logarithmic slowing down of diffusion.\bl
Strong evidence that the bias drives the particle into the deepest dangling 
ends can also 
 be drawn by looking at the amplitude dependence of occupation 
probabilities. Indeed 
 we notice that the $P(t)$ of tip sites which are at 
maximum chemical distance from
 the bias source, have an amplitude  factor proportional to $1/ 2^n$ (see 
 eq.(4.3c); an identical result holds for site 6). The number of such sites
 grows
 just as $2^n$ with the order $n$ of the T fractal, because the branch 
containing 
 the absorbing site has not to be taken into account due to its depleting 
nature 
 (compare for example eq.(4.3c) with (4.3d) and (4.5)). Hence we infer 
that, after 
 a transient time, probability accumulates into these deep dangling ends, 
 redistributing then slowly against the biasing force to sites with 
higher coordination 
 and/or in less favourable positions. Amplitude factors for the 
probability of every other 
 kind of site are in fact proportional to $\d W^{\eta} / 2^n$, 
where the exponent
 $\eta$ depends manifestly, though non trivially, on the chemical 
distance between
 the site under consideration and those deepest dangling end, namely 
on the number of 
 unfavourably biased bonds one has to cross going from the latter to
 the former. 
 In an infinite size limit, occupation probabilities of sites with coordination 
 $>1$ will then be vanishingly small compared with those relative to the 
dangling ends, in 
agreement with our hypothesis.\bl
Finally we would like to point out the peculiar features of the sink site. 
This site can not be considered, like other tips, 
as a probability accumulating point, of course, and 
 the same applies to its neighbouring sites (e.g. number 9 in Fig.2.1b).
 Such circumstance reflects on 
the $W^{2^n}$ and $W^{2^{n-1}}$ prefactors in eqs.(4.3d) and
 (4.5), respectively. Moreover $\pt_{10}$ (i.e. $\pt_{FP}$) has actually two 
simple poles:
 $\om_{c{\scriptscriptstyle 1}} = -{1\over \wp}$ and $\om_{c{\scriptscriptstyle
 2}} {\buildrel {
\scriptscriptstyle
n\tende +\infty}\over \cong} -{1 /\wp t_{n}^* }$, as 
 it appears from eqs.(3.6,12). Therefore two characteristic times exist, 
the first
 deriving from $\om_{c{\scriptscriptstyle 1}}$, very short and connected to 
the transient time when 
 the initial flux of probability entering the side branch is immediately 
absorbed
 by the sink. The second, $t_{n}^*$, is quite long and corresponds to the 
asymptotic 
regime when probability formerly accumulated  into deep dangling ends 
slowly reflows 
in the whole lattice and then is absorbed by the sink.\bl
To probe a possible dependence of diffusion process on initial conditions,
 we have 
studied the same model choosing different input sites, namely no longer
  site 1, 
but either 4 or 7 of Fig.2.1b, successively. We arrived at exactly the 
same results for 
$\lb t\rb^{(n)}$, $t_{n}^*$, $\pfp (t)$ and occupation probabilities. 
Hence we claim 
that the above results and considerations hold for our model 
independently of initial 
conditions.\bl
The description of the dynamic model we are concerned with, should be
independent on
 the particular realization of the bias pattern. To get convinced of this, a 
less asymmetric situation 
 than that presented in Fig.2.1, has also been considered. Again, 
starting from 
a T fractal, we suppose to orient the arrows of topological bias 
as in Fig.4.1.
 This new choice, 
together with the conventions on $W_{ij}$ specified in section 2 [21], implies 
that the bias 
pushes  a diffusing particle  away from the T fractal central site (number 2 in
 Fig.4.1a, or 4 in Fig.4.1b). In addition the three tip-sites 
(e.g. 1,7,10 in the 
2-order lattice of Fig.4.1b) are assumed to be absorbing, and the central one to
be the input point of the walker. Master equations and their Laplace 
transforms are equivalent
to (2.1) and (2.4) respectively, with the difference that now $\alpha_0 (\om )$ 
refers to the central site, which has three nn, but is not dynamically 
equivalent 
 to other 3-coordinated sites. The method outlined in section 3 and the 
 present one applies of course to the latter model too. Considering
 an $n$th-order tree one finds, for relevant quantities:
$$\d \lb t\rb^{(n)} \propto {L\over 2W^{L/2}} \eqno(4.7a)$$
$$\d t_{n}^* \simeq {(1+W)\over \wp }{2^n \over W^{2^{n-1}}} \eqno(4.7b)$$
$$\d P_{1,7,10} (t) \approx {W^{2^{n-1}} \over 3\  2^{n-1} +3\wp W^{2^{n-1}}} 
e^{-t/t_{n}^*} \eqno(4.7c)$$
$$\d P_4 (t) \approx {(1-W^{2^n})\over 1-W^2} {W^{2^{n}} \over 3\  2^n } e^{-1/t_{n}^*}
\eqno(4.7d)$$
$$\d P_{3,6,9}(t) \approx {(1-W^{2^{n-1}}) \over 1-W^2} {1\over 3\ 2^n} 
\d e^{-t/t_{n}^*} \eqno(4.7e)$$
$$\d P_{2,5,9}(t) \approx {(1-W^{2^{n-1}}) \over 1-W^2} {2W^{2^{n-1}}\over 3\  2^n} 
e^{-t/t_{n}^*} \ \ \ .\eqno(4.7f)$$
Site indices, shown in Fig.4.1b, are those which would apply after $n-1$ 
decimations. Taking into account that the linear size of the $n$th-order 
lattice is $2^{n+1}$, but that each of its three branches is indeed equivalent 
to the $(n-1)-$order lattice of Fig.2.1, and its length is thus $L=2^n$, 
then results (4.7) are totally consistent with previous ones. This is not 
surprising in view of the threefold symmetry implied by 
the conditions we imposed.\bl
So, our analysis 
applies to different realizations of the present model. It can also be shown
 that our approach can be extended 
to a wider family of ramified fractals, like for example that in Fig.2.2, in 
$d\ge 2$ dimensions.\par
{\bf 5 Conclusions}\bl
We have considered the problem of random walks on deterministic self-similar 
trees in the presence of a topological biasing field. Relying on FSS ideas,
 we have developed a method 
 to extract 
the FSS behaviour of the average FPT, $\lb t\rb$, and characteristic time, 
$t^*$. The method is phenomenological and is based on successive decimations and
 rescalings of the set of master equations for a finite $n$th-order lattice.
 $\lb t\rb$ and $t^*$ are found to scale 
in the same way with the system size, $L$, i.e. 
$\lb t\rb\sim t^* \sim \d{L\over W^{L/2}}$. Thus, dynamics is controlled 
by a unique time scale 
growing exponentially with the tree's size. Since infinite fractal trees 
contain branches of any length scale, where the particle can be trapped in, 
we argue that the resulting slowing down should indeed lead to the logarithmic 
diffusion 
conjectured for infinite systems. Moreover, the existence of a unique
 time scale 
reveals a substantial equivalence between the fixed time and the fixed point of
 observation ensembles, governing the behaviour of 
$\lb R(t)\rb$ and $\lb t\rb$, 
respectively. The scaling of $t^*$ implies also a divergence of the 
dynamical exponent $z$ and thus a singularity in standard RG 
transformations for infinite lattice. The 
possibility 
to carry on a phenomenological RG strategy in the presence of singular 
transformations, via the 
expedient of considering systems of finite size, should be seen as the
main achievement 
of our technique, in view of the serious complications involved
in the RG analysis of infinite systems [12]. It is also worth noting that we 
produced here an interesting example of the validity of the FSS hypothesis, 
notwithstanding the existence of a singularity in the RG group.\bl
Finite size analysis has been pushed further here to calculate the long time 
behaviours 
of FPP and occupation probabilities, by combining asymptotic expansion of the 
respective 
Laplace transforms with RG rescalings. All probabilities are found to decay 
exponentially at long $t$ , again with the same characteristic time 
$t^*$, as expected. 
This fact, of course, supplies stronger confidence to the picture of
 logarithmic diffusion 
generated by the mechanism of bias induced trapping into dangling 
ends. Amplitude 
factors of occupation probabilities offer further insight
 into this dynamic process. Indeed, the occupation probabilities of sites 
located in 
the deepest dangling ends have amplitude proportional to $1/2^n$, which is just 
the inverse of their number in an $n$th order lattice. Moreover,
 probabilities relative 
to sites with higher coordination or in a less favourable position with respect 
to the biasing field, have amplitudes proportional to $W^{\eta}/2^n$ ($W<1$)
, with $\eta$ depending on the chemical distance 
between the deepest dangling ends and the site under consideration. Thus 
probability first accumulates into the longest branches of the fractal 
under the bias
 drift; then it very slowly reflows against the 
bias to the whole lattice, generating 
the logarithmic slowing down of diffusion.\bl
The above results have been demonstrated to 
hold in general for the system under 
discussion, no matter which initial and boundary 
conditions are chosen, and independently 
of the particular realization of the model. Our methods can be 
applied to a wide 
variety of finitely ramified fractals [18], for which they provide the same 
picture of logarithmic biased diffusion outlined here.\par
{\bf Acknowledgments}\bl
We are indebted to A. Maritan for collaboration in the early
 stages of our work on logarithmic diffusion.
\vfill\supereject
\centerline{\bf References}\bl 
{\item{[1]} E. Marinari, G. Parisi, D. Ruelle and P. Widney, 
{\it Phys. Rev. Lett. \/}{\bf 50}, 1223 (1983).}
\item{[2]} D.S. Fischer, {\it Phys. Rev. A \/}{\bf 30}, 960 (1984); D.S. 
Fisher, D. Friedan, Z. Qiu, S.J. Shenker and S.H. Shenker,
 {\it Phys. Rev. A \/}{\bf 31}, 3841 (1985).
\ref{3}{R. Durrett}{Commun. Math. Phys.}{104}{87}{1986} 
\ref{4}{R.L. Bulmberg Selinger, S. Havlin, F. Leyvraz, M. Schwartz and H.E. 
Stanley}{Phys. Rev. A}{40}{6755}{1989}
{\item{[5]} M. Nauenberg, {\it J. Stat. Phys. \/}{\bf 41}, 803 (1985).} 
{\item{[6]} A. Bunde, S. Havlin, H.E. Roman, G. Schildt and H.E. Stanley, 
{\it J. Stat. Phys. \/}{\bf 50}, 1271 (1988).}
\ref{7}{Ya.G. Sinai}{Theory Prob. its Appl.}{27}{256}{1982}
{\item{[8]} D. Stauffer, {\it J. Phys. A \/}{\bf 18}, 1827 (1985).}
\item{[9]} A. Bunde, H. Harder, S. Havlin and H.E. Roman, 
{\it J. Phys. A \/}{\bf 20}, L865 (1987).
\ref{10}{S. Havlin, A. Bunde, Y. Glaser and H.E. Stanley}
{Phys. Rev. A}{34}{3492}{1986}
{\item{[11]} Sinai's and related results [1,3-7] are intuitively  explained
 once it is realized that over a distance $R$ along the 1D chain, a potential 
barrier $\propto \sqrt{R}$ develops by adding local random biases with zero 
average. On the basis of Arrhenius law, a time $t\propto\d e^{\sqrt{R}}$ is 
needed to overcome such a barrier, then $R\sim \ln^2 t$ follows.}
\ref{12}{A. Maritan, G. Sartoni and A.L. Stella}{Phys. Rev. Lett.}{71}{1027}
{1993}
\ref{13}{S. Havlin and D. ben-Avraham}{Adv. Phys.}{35}{695}{1987}
\ref{14}{A. Giacometti, A. Maritan and A.L. Stella}{Int. J. Mod. Phys. B}
{5}{709}{1991}
\ref{15}{R. Stinchcombe}{J. Phys. A}{18}{L591}{1985}
\ref{16}{M. Suzuki}{Prog. Theor. Phys.}{58}{1142}{1977}
\ref{17}{B. Kahng and S. Redner}{J. Phys. A}{22}{887}{1989}
{\item{[18]} This is the most simple example in 
$d=2$ of a whole class of ramified fractals 
in $d$ dimensions, which can all be treated with our methods (see Fig.2.2).}
\ref{19}{A. Giacometti, A. Maritan and A.L. Stella}{Phys. Rev. B}{38}{2758}
{1988}
{\item{[20]} It is important that $\om_c {\buildrel 
{\scriptscriptstyle n\tende +\infty}\over \ltende}0$, because 
we look at the asymptotic behaviour, $\om \tende 0$, 
of $\pt_{FP}$. Were the simple pole located elsewhere, we could not find it by 
the asymptotic analysis.}
{\item{[21]} Actually the condition $3\wp \leq 1$ has to be added to what 
stated in 
section 2, to take into account the peculiar nature of the central site.}
\vfill\supereject
\centerline{FIGURES}\bl
\item{FIGURE 2.1} The first three successive iterations of the 
T fractal, (a), (b) 
and (c) respectively. The arrows on each bond indicate the direction  of the 
topological bias.\bl
\vskip 1.5truecm\noindent
\item{FIGURE 2.2} Example of T fractal with coordination 4 or 1, in $d=2$. 
Analogous lattices can be drawn in $d>2$.\bl
\vskip 1.5truecm\noindent
\item{FIGURE 4.1} The first two iterations of the symmetric T fractal lattice, 
(a) and (b) respectively, with a topological bias on each bond.\bl
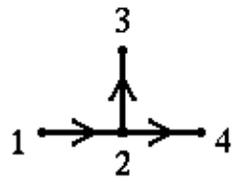
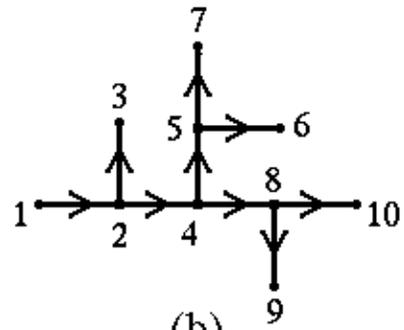
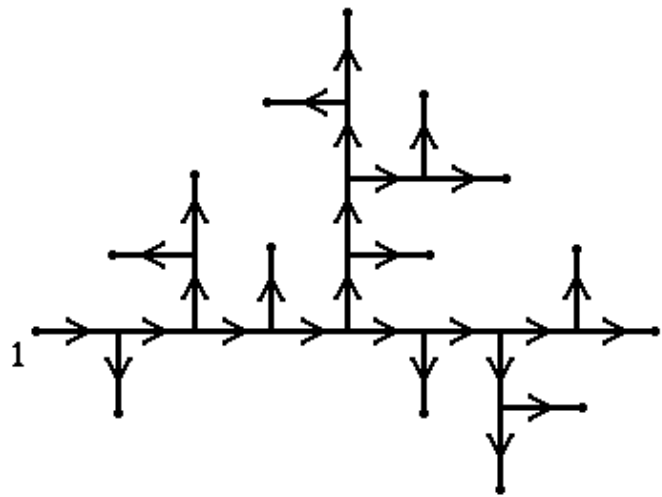
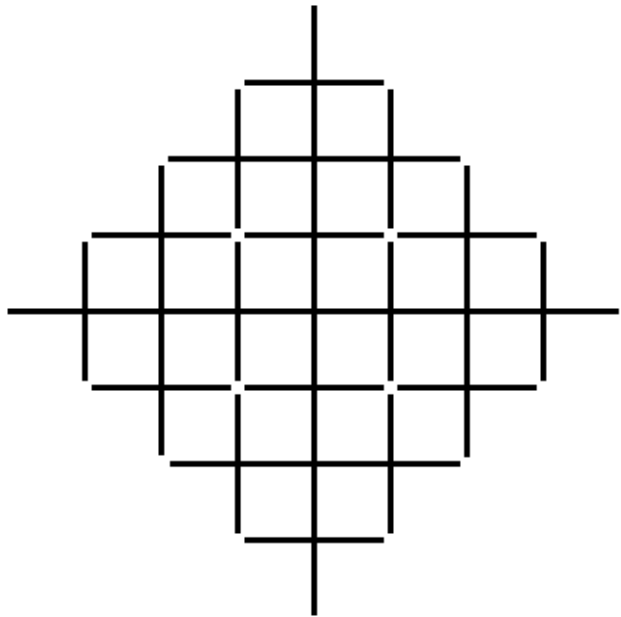
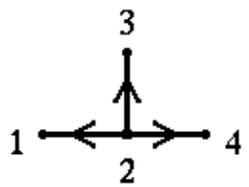
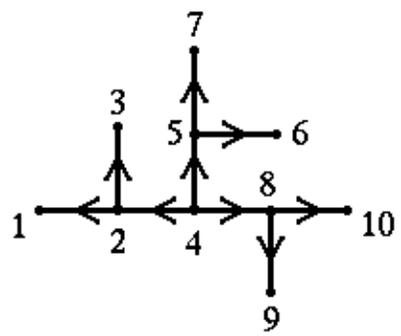
\bye